\documentstyle [12pt] {article}

\def\L{{\cal L}}
\def\R{{\rm I \!\!\, R}}
\def\be{\begin{equation}}
\def\ee{\end{equation}}
\def\bea{\begin{eqnarray}}
\def\eea{\end{eqnarray}}

\begin{document}
\title{ {\bf Homothetic perfect fluid  space-times}}

\author{\normalsize J.Carot\thanks{{\sl Email:}dfsjcg0@ps.uib.es}
\hspace{1.0cm} A.M. Sintes\thanks{{\sl Email:}dfsaso4@ps.uib.es}\\
\normalsize Departament de  F\'{\i}sica, Universitat de les Illes Balears\\
\normalsize E-07071 Palma de Mallorca, SPAIN}
\date{}
\maketitle

\begin{abstract}
A brief  summary of results on homotheties in General Relativity is given,
including  general
information about space-times admitting an $r$-parameter group of homothetic
transformations for $r>2$, as well as some specific results on perfect fluids.
Attention  is then focussed on  inhomogeneous models, in particular on those
 with  a homothetic group $H_4$ (acting multiply transitively) and $H_3$. A
classification of all possible Lie algebra structures along with (local)
coordinate expressions for the metric and homothetic vectors is then provided 
(irrespectively of the matter content), and  some new perfect fluid  solutions
are given and briefly discussed.

 \end{abstract}

\section{Introduction}

This paper is devoted to the study of space-times admitting an intransitive
group of homotheties, with a view towards those which can be interpreted as
perfect fluid solutions of Einstein's field equations \cite{Kramer}.

A collection of important results regarding generic properties of
space-times admitting homothetic transformations can be found in 
\cite{Hall88}-\cite{HallCos88} (and references cited therein), and in   
\cite{Hall90} where the case of multiply transitive action is thoroughly
studied by Hall and Steele.

The study of this subject began with  the
pioneering paper by Cahill and Taub \cite{CaT}, followed by the works of
Eardley \cite{Eardley,Eardley2}. From then on, homotheties have been 
studied in connection with a wealth of situations of physical interest in
classical general relativity as well as in cosmology, see  \cite{p1,p2,p3}
for interesting reviews on homothetic solutions.


The paper is organized as follows: section 2 contains a brief summary of
results on groups of homotheties and space-times admitting them,  the
implications that they have on perfect fluids and the
physical quantities characterizing them (density, pressure, velocity,...), 
and we also summarize all the general information about space-times
admitting an $r$-parameter group of homothetic transformations for $r>2$.
This includes: dimension of the homothetic and isometric algebras as well as
that of the orbits they act on respectively, together with its nature
(spacelike, timelike or null), allowed Petrov and Segre types of the Weyl and
Ricci tensors, and whether perfect fluid solutions exist or not.  Most of the
contents of this section is a review of dispersed results in the literature,
but we considered useful to gather them all in a single table. Some of the
results are, as far as we are aware of, new (see especially those cases where
null orbits occur).

In sections 3 and 4  a
classification of all possible Lie algebra structures along with (local)
coordinate expressions for the metric and homothetic vectors, are given
for the cases $H_4$ (acting multiply transitively on three-dimensional
orbits, in keeping with our assumption of intransitive action) and $H_3$
respectively. 
These classes of space-times can be understood as generalizations to the
Kantowski-Sachs and Bianchi models respectively. The characterizations
provided are independent of the field equations, and therefore they may have
applications other than those considered here (perfect fluids).

In particular, section 3 contains some review material on the case $r=4$,
together with some new results, all of them presented in a unified manner,
extending the
work of Wu \cite{Wu}, Cahill and Taub \cite{CaT} and Shikin \cite{Shikin}.
The general perfect fluid solution is then given in
certain, well-defined and invariantly characterized subcases. Whenever this
is not possible, a few selected examples are presented. It is
assumed that the matter satisfies the weak and dominant energy conditions,
and expressions for the
 kinematical quantities  (acceleration,
expansion, deceleration parameter, shear and vorticity) are provided for each
case.

Finally, section 4 contains some new  solutions for the case $r=3$ and
appropriate references to related work on  this issue.
We summarize
the results concerning the topology of the Killing orbits and the Bianchi
classification of the homothetic algebras. We distinguish the cases where
the Killing subalgebra is Abelian from that where it is non-Abelian.  
Attention is then  restricted to the orthogonally transitive case, giving for
each possible Lie algebra structure the coordinate forms of the proper
homothetic vector  field  and the metric. In the Abelian case we distinguish
three different classes  of such models, depending on the orientation of the
fluid flow relative  to the homothetic orbits. The case in which we are more
interested is the so-called \lq\lq tilted" where new solutions are found.
Finally, we provide  explicit forms for the homothetic vector field
and the metric in the case of a (maximal)   non-Abelian $G_2$, and, although
no perfect fluid solutions have been found, we briefly discuss some
properties.

\section{Basic facts about homotheties}
\subsection{Definition and properties}

Throughout this paper $(M,g)$ will denote a space-time: $M$ then being a
Hausdorff, simply connected, four-dimensional manifold, and $g$ a Lorentz
metric of signature (-,+,+,+). All the structures will be assumed smooth.

A global vector field $X$ on $M$ is called homothetic if either one of the
following equivalent conditions holds on a local chart
\be
\L_Xg_{ab}=2 n g_{ab} \ ,\ \  \ X_{a;b}=  n g_{ab}+F_{ab}\ ,
\label{r1}
\ee
where $n$ is a constant on $M$ , $\L$ stands for the Lie derivative
operator, a semi-colon denotes a covariant derivative with respect to
the metric connection, and $F_{ab}=-F_{ba}$ is the so-called homothetic
bivector. If $n \not=0$, $X$ is called proper homothetic and if
$n=0$, $X$ is a Killing vector (KV) on $M$. For a geometrical
interpretation of (\ref{r1}) we refer the reader to \cite{Hall88,Hall90b}.

A necessary condition that $X$ be homothetic is
 \be
{X^a}_{;bc}={R^a}_{bcd}X^d\ ,
\label{r2}
\ee
where ${R^a}_{bcd}$ are the components of the Riemann tensor in the above
chart; thus, a homothetic vector field (HVF) is a particular case of affine
collineation \cite{HallCos88} and therefore it will satisfy \be
\L_X {R^a}_{bcd}=\L_X R_{ab}=\L_X{C^a}_{bcd}=0 \ ,\label{r3}
\ee
where $R_{ab}$ ($\equiv {R^c}_{acb}$) and ${C^a}_{bcd}$ stand, respectively
for the components of the Ricci and the conformal Weyl tensor.

The set of all  HVFs on $M$ forms a finite
dimensional Lie algebra under the usual bracket operation and will be
referred to as the homothetic algebra, ${\cal H}_r$, $r$ being its dimension.
The set of all KVs  on $M$ also forms a finite dimensional Lie
algebra,  the Lie algebra of isometries, which will be denoted as
 ${\cal G}_s$ ($s$ being its dimension), and   one has that ${\cal G}_s
\subseteq {\cal H}_r$ (i.e.,  ${\cal G}_s$ is a subalgebra of ${\cal H}_r$).
Furthermore, it is immediate to see by direct computation that the Lie bracket
of an HVF with a KV is always a KV and that, given any two proper HVFs, there
always exists a linear combination of them which is a KV. From these
considerations it immediately follows that the highest possible dimension of
${\cal H}_r$ in a four-dimensional manifold is $r=11 $.

If $r \neq s$ then $s=r-1$ necessarily, and one may choose a basis
$\{ X_1,\cdots, X_{r-1},X\} \equiv \{X_A\}_{A=1\cdots r}$ for ${\cal H}_r$,
in such a way that $X$ is proper homothetic and $X_1, \cdots,X_{r-1}$ are
Killing vector fields spanning ${\cal G}_{r-1}$. If these vector fields in
the basis of ${\cal H}_r$ are all complete vector fields, then each member
of ${\cal H}_r$ is complete and Palais' theorem
\cite{Hall88b,Palais57,Brickell70}
guarantees the existence of an $r$-dimensional Lie group of
homothetic transformations of $M$ ($H_r$) in a well-known way; otherwise,
it gives rise to a local group of local homothetic transformations of $M$
and, although the usual concepts of isotropy and orbits still hold, a little
more care is required \cite{Hall90}.

The following result \cite{Hall90,Bilyalov} will be useful: 

The orbits
associated with ${\cal H}_r$ and ${\cal G}_{r-1}$ can only coincide if
either they are four-dimensional or three-dimensional and null. (This
result still holds if ${\cal H}_r$ is replaced by the
conformal Lie algebra ${\cal C}_r$ and does not depend on the maximality of
${\cal H}_r$ or ${\cal C}_r$).

The set of zeroes of a proper HVF, i.e.,  $\{p\in M : X(p)=0\}$ (fixed points
of the homothety), either consists of topologically isolated points, or else
is part of a null geodesic. The latter case corresponds to the well-known
(conformally flat or Petrov type N) plane waves \cite{Hall88,Alex85}.

At any zero of a proper HVF on $M$ all Ricci and Weyl eigenvalues must
necessarily vanish and thus the Ricci tensor is either zero or has Segre
type $\{(2,11)\}$ or $\{ (3,1)\}$ (both with zero eigenvalue), whereas the
Weyl tensor is of the Petrov type $O$, $N$ or $III$ \cite{Hall88}
(see also \cite{Beem} for vacuum space-times).

\subsection{Perfect fluids}
The energy-momentum tensor of a perfect fluid is given by
\be
T_{ab}=(\mu+p)u_au_b+pg_{ab}\ ,
\label{r4}
\ee
where $\mu$ and $p$ are, respectively, the energy density and the pressure
as measured by an observer comoving with the fluid, and $u^a$ ($u^au_a=-1$)
is the four-velocity of the fluid. If $X$ is an HVF then, from Einstein's
Field Equations (EFE) it follows that
\be
\L_XT_{ab}=0\ ,
\label{r5}
\ee
and this implies in turn \cite{Eardley}
\be
\L_Xu_a=n u_a\ , \quad \L_Xp=-2n p\ , \quad \L_X\mu=-2n
\mu\ . \label{r6}
\ee
Thus, the Lie derivatives of $u_a$, $p$ and $\mu$ with respect to a KV
vanish identically.

If a barotropic equation of state exists, $p=p(\mu)$,
and the space-time admits a proper HVF $X$ then \cite{Wainwright}
\be
p=(\gamma-1)\mu\ ,
\label{r7}
\ee
where $\gamma$ is a constant ($0\le\gamma\le 2$ in order to comply with
the weak and dominant energy conditions). Of particular interest are the
values $\gamma=1$
(pressure-free matter, \lq\lq dust") and $\gamma=4/3$ (radiation fluid).
In addition, the value $\gamma=2$ (stiff-matter) has been considered in
connection with the early Universe. Furthermore, values of $\gamma$
satisfying $0\le\gamma<2/3$, while physically unrealistic as regards a
classical fluid, are of interest in connection with inflationary models of
the Universe. In particular, the value $\gamma=0$, for which the fluid can
be interpreted as a positive cosmological constant, corresponds to
exponential inflation, while the values $0<\gamma<2/3$ correspond to power
law inflation in FRW models \cite{Barrow86}, but it is customary to
restrict $\gamma$ to the range $1\le\gamma\le 2$.

If the proper HVF $X$ and the four-velocity $u$ are mutually orthogonal
(i.e., $u^aX_a=0$) and a barotropic equation of state is assumed, it follows
that $\gamma=2$, i.e., $p=\mu$ stiff-matter \cite{Eardley}, on the other
hand, if $X^a=\alpha u^a$ the fluid is  then shear-free. Further information
on this topic can be found in \cite{McInt76,McInt78,McInt79}.

\subsection{The \lq\lq dimensional count-down"}
In this subsection, the maximal Lie algebra of global HVF on $M$ will be
denoted as ${\cal H}_r$ ($r$ being its dimension), and it will be assumed
that at least one member of it is proper homothetic.

The case of multiply transitive action is thoroughly studied in
\cite{Hall90}, and we shall refer the reader there for details;
nevertheless, and for the sake of completeness, we summarize in the following
table the results given there, which follow invariably from considerations on
the associated Killing subalgebra and the fixed point structure of the proper
HVF.
Futhermore, we have added a few other results, also in the literature or
following straightforwardly from those, so as to complete the study down to
dimension 4.

{\small
\begin{center}
\begin{tabular}{|r|c|c|c|c|c|c|} \hline
$r$ & $O_m$ & $K_n$ & {\it Petrov} & {\it Segre} &{\it Interpretation} &
{\it PF, info.} \\ \hline

$11$ & $M$ & $M$ & $O$ & $0$ & Flat & $\not\exists $ \\
$10$ & $M$ & $M$ & - & - & Not Possible & $\not\exists $ \\
$9$  & $M$ & $M$ & - & - & Not Possible & $\not\exists $ \\
$8$  & $M$ & $M$ & $O$ & $\{(2,11)\}$ & Gen. Plane wave & $\not\exists $ \\
$7$  & $M$ & $M$ & $N$ & $0,\{(2,11)\}$ & Gen. Plane wave & $\not\exists $ \\
$7$  & $M$ & $T_3$ & $O$ & $\{(1,11)1\}$ & Tachyonic Fluid & $\not\exists $
\\
$7$  & $M$ & $N_3$ & - & - & Not Possible & $\not\exists $ \\
$7$  & $N_3$ & $N_3$ & $O$ & $\{(2,11)\}$ & Gen. Plane wave & $\not\exists $
 \\
$7$  & $M$ & $S_3$ & $O$ & $\{1,(111)\}$ &Perfect Fluid & FRW \\
$6$  & $M$ & $M$ & - & - & Not Possible & $\not\exists $ \\
$6$  & $N_3$ & $N_3$ & $N$ & $\{(2,11)\}$ & Gen. Plane wave & $\not\exists $
 \\
$5$  & $M$ & $M$ & - & - & Not Possible & $\not\exists $ \\
$5$  & $M$ & $N_3$ & - & - & - & $\not\exists $ \\
$5$  & $N_3$ & $N_3$ & - & - & Not Possible & $\not\exists $ \\
$5$  & $M$ & $T_3$ & $D,N,O$ & $\{1,1(11)\},\{2,(11)\}$  &  & LRS \\
$5$  & $M$ & $S_3$ & $D,O$ & $\{(1,1)11\},\{(2,1)1\}$  &  & LRS \\
$4$  & $M$ & $N_3$ & $II, III, D,N,O$ & $\{(1,1)(11)\},\{(2,11)\}$ & Plane
waves  & $\not\exists$ \\
$4$  & $N_3$ & $N_3$ & - & - & Not Possible & $\not\exists $ \\
$4$  & $M$ & $T_3$ &  &  &  & Bianchi \\

$4$  & $M$ & $S_3$ &  &  &  & Bianchi \\
$4$  & $O_3$ & $N_2$ & $N,O$  & $\{3,1 \}, \{2,11\},\{(1,1)11\}$ &  &
$\not\exists$ \\
$4$  & $O_3$ & $T_2$ & $D,O$  & $\{(1,1)11\}$ &  & $\not\exists$ \\

$4$  & $O_3$ & $S_2$ & $D,O$  & $\{ - (11) \}$  &  & $\exists$
\\         \hline
\end{tabular}
\end{center}

\begin{center}
{\bf Table 2.1}
\end{center}
}
The first entry in the table gives the dimension of the group of
homotheties, the second and third entries stand for the nature and dimension
of the homothetic and Killing orbits respectively (e.g.: $N_2$, $T_2$ and
$S_2$ denote Null, Timelike and Spacelike two-dimensional orbits respectively,
$O_3$ stands for three-dimensional orbits of either nature, timelike,
spacelike or
null), the fourth and fifth entries give the Petrov and Segre type(s) of
the associated Weyl and Ricci tensors (in the latter case it is to be
understood that all possible degeneracies of the given types, can in
principle occur, including vacuum when possible). Finally, the last two
entries give respectively the possible interpretation whenever it is in
some sense unique, and the existence or non-existence of perfect fluid
solutions for that particular case, along with some supplementary
information; thus FRW stands for Friedmann-Robertson-Walker, LRS for
Locally Rotationally Symmetric, and Bianchi refers to that family of
perfect fluid solutions. The cases that cannot arise are labeled as \lq\lq
Not Possible", and wherever no
information is given on the Petrov and Segre types, it is to be understood
that all types are possible in principle. The Segre type of the Ricci tensor
of the case described in the last row, is unrestricted except in that it
must necessarily have two equal (spacelike) eigenvalues; perfect fluid
solutions of these characteristics constitute special cases of spherically,
plane or hyperbolically symmetric perfect fluid space-times. For further
information
on LRS space-times, see \cite{Ellis67,Stewart68}; for the case $r=4$ transitive
and null three-dimensional Killing orbits, see \cite{Kramer,Rosq}. Regarding
spatially homogeneous Bianchi models, see \cite{Rosq,Ryan,Hsu1,Hsu2}; and for
the last three cases occurring in the table, see respectively
\cite{Barnes79,Goenner70},
\cite{Kramer,Goenner70}, and \cite{Kramer}.

The case $r=3$ has an associated Killing subalgebra ${\cal G}_2$ and the
respective dimensions of their orbits are 3 and 2 (see for instance
\cite{Ali1,Uggla92,Mars,Hewitt91} and references cited therein).
When the Killing subalgebra has null orbits, the metric is of Kundt's class
\cite{Kundt61} and perfect fluids are excluded. If the Killing orbits are
timelike, the solutions can then be interpreted as special
cases of axisymmetric stationary space-times (provided that regularity
conditions hold on the axis \cite{Kramer,Mars}), and if they are spacelike
as special cases of inhomogeneous cosmological solutions or cylindrically
symmetric space-times. In both cases, perfect fluid solutions have
been found.

\section{The $H_4$ case} 
The associated isometric group for perfect fluid space-times acts necessarily on
$T_3$, $S_3$ or $S_2$ orbits (see {\bf Table 2.1}). In the 
intransitive case,
the $G_3$ must act multiply transitively on two-dimensional surfaces of maximal
symmetry $S_2$, which are then
of constant (positive, zero or negative) curvature and admit
orthogonal surfaces  \cite{Schmidt}.

Possibly  no problem in this context has been more exhaustively studied
than that  of  spherically symmetric homothetic space-times, beginning with the
seminal paper of Cahill and Taub \cite{CaT} and continuing with recent papers 
by  Ori and  Piran \cite{Ori2},
 Carr and  Yahil \cite{Yahil},
 Henriksen and  Patel \cite{Patel}, and Foglizzo and Henriksen \cite{Fog} among
others (see references therein). Homothetic space-times with plane symmetry
are also considered by Shikin \cite{Shikin}.

What we attempt in this section,
rather than presenting a survey of the models existing in the literature,
 is to study in a unified manner all possible
cases, i.e., homothetic orbits of either nature (timelike, spacelike or  null
at every point $p\in M$ and the more general case in which their nature
varies from point to point) as well as the different possibilities for the
curvature $k$  of the isometry orbits (spherical and plane symmetry as well 
as the $k=-1$ case); thus, extending previous works by Wu
\cite{Wu}, where only spacelike homothetic orbits are considered (i.e., type
$B$ and some type $C$ solutions in our classification below), and by  Cahill
and Taub \cite{CaT}, and  Shikin \cite{Shikin} where only spherical and plane
symmetry are considered respectively. 

As it is well known, the space-time metric
can be written as \cite{Kramer}:
\be
ds^2=-A^2(r,t)dt^2+B^2(r,t)dr^2+F^2(r,t)(dy^2+\Sigma^2(y,k)dz^2)\ ,
\label{r99} \ee

\be
\Sigma(y,k)=\left\{ \begin{array}{ll}
\sin y & k =+1 \\
y & k =0 \\
\sinh y & k =-1 \ .
\end{array}
\right. 
\ee
The Killing vectors being $\xi_1 = \sin z \partial_y +{\Sigma'\over \Sigma}\cos
z\partial_z $, $\xi_2=  \cos z \partial_y -{\Sigma'\over \Sigma}\sin z
\partial_z$, and $\xi_3= \partial_z$, where
the dash denotes a derivative with respect to $y$, and they satisfy the
following commutation relations
\be
[\xi_1, \xi_2]=k\xi_3\ , \quad [\xi_2, \xi_3]=\xi_1\ , \quad [\xi_3, \xi_1]=
\xi_2 \ .\ee

Assuming the existence of a proper
HVF, $X$, and since its commutator with a KV must be a KV, the
Jacobi identities imply the following structures for ${\cal H}_4$
\bea
&(I)&[X,\xi_i]=0\ , \quad i=1,2,3 \ ,\quad k=0,\pm 1 \ , \\ & &
\quad X=X^t(t,r)\partial_t+X^r(t,r)\partial_r\ ,  \\
&(II)&[X,\xi_1]=\xi_1\ , \quad
[X,\xi_2]=\xi_2 \ ,\quad [X,\xi_3]=0 \ ,\quad k=0 \ , \\ & &
\quad X=X^t(t,r)\partial_t+X^r(t,r)\partial_r-y\partial_y \ .\eea

Requiring the coordinate system to be a comoving one (i.e., $u_a=-A(t,r)
{\delta}^t_a$)
the HVF takes the form
\be
X=X^t(t)\partial_t+X^r(r)\partial_r+X^y(y)\partial_y\equiv \hat X+
X^y(y)\partial_y\ ,
\ee
and the following possibilities then arise:
\bea
&(A)&\hat X=\partial_t \ ,\label{r100}\\
&(B)&\hat X=\partial_r \ ,\label{r101}\\
&(C)&\hat X=\partial_t+\partial_r \ .\label{r102} \eea
The form ($A$) corresponds to $u$ being tangent to the timelike homothetic
orbits, ${\cal T}_3$, ($B$) corresponds to the case of spacelike homothetic
orbits, ${\cal S}_3$, orthogonal to the fluid flow, and ($C$) is the most
general (tilted) case, including also the possibility of having null
homothetic orbits, ${\cal N}_3$, when the functions $A(t,r)$ and $B(t,r)$ in
(\ref{r99}) equal each other.

The homothetic equation (\ref{r1})
specialized
to
the metric (\ref{r99}) yields then the following possibilities:
\hfill\break

\noindent {\bf Case ($A$)}

\begin{center}
\begin{tabular}{|l|c|c|c|c|c|}\hline
$A$ & $k$ & $X$ & $A^2(t,r)$ & $B^2(t,r)$ & $F^2(t,r)$\\\hline
$I$ & $-1,0,1$ & $\partial_t$ & $e^{2t}H^2(r)$ & $e^{2t}H^2(r)$ &
$e^{2t}f^2(r)$ \\ \hline
$II$ & $0$ & $\partial_t-y\partial_y$ & $e^{2nt}H^2(r)$ & $e^{2nt}H^2(r)$ &
$e^{2(n+1)t}f^2(r)$ \\ \hline
\end{tabular}
\end{center}
\begin{center}
{\bf Table 3.1}
\end{center}

\noindent where $t$ has been scaled in  $AI$ so as to make $n=1$ in
(\ref{r1}). Solving now the field equations for a perfect fluid in each one
of the two above cases, we find that the only possible solution for the family
$AI$ has an equation of state $\mu +3p=0$ and it admits (at least) one
further Killing vector,
 thus being a particular case of  ${\cal H}_5$ and we shall not study it here.

For $AII$, two families arise which depend on the value of $n$:

\[
\begin{array}{|c|c|c|c|c|c|}\hline
AII & H(r) & f(r) & \mu & \gamma  \\\hline\hline
\begin{array}{c} n\in(-\infty,-3)\cup  \\ (-2,-1)\cup (0,+\infty)
\end{array}  & a(\cosh (\alpha r))^{-{1 \over \alpha}} & ({\displaystyle H
\over\displaystyle a})^{n+1}
 &
{\displaystyle \beta \over\displaystyle a^{2\alpha}} e^{-2nt}H^{2(\alpha-1)} 
& {\displaystyle 2n \over \displaystyle 2+3n}  \\ \hline n\in(-3,-2) & 
a(\sinh (\alpha r))^{-{1 \over \alpha}} &  ({\displaystyle H
\over\displaystyle a})^{n+1} & -{\displaystyle\beta \over\displaystyle
a^{2\alpha}} e^{-2nt}H^{2(\alpha-1)} & {\displaystyle 2n \over \displaystyle
2+3n} \\ \hline \end{array} \]
\begin{center}
{\bf Table 3.2}
\end{center}

\noindent where 
\be\alpha = {2(n+1) \over n+2} \ , \quad\beta =
{(n+1)(n+3)(3n+2) \over n^2(n+2)} \ ,
\ee
and 
$a$ is a constant. The vorticity is zero and the volume expansion $\theta$,
deceleration parameter $q\equiv-1-3\dot\theta /\theta^2$, acceleration
$\dot u_a$, and non-vanishing shear tensor components $\sigma_{ab}$ can be
given as:
\be
\theta={2+3n \over H}e^{-nt}\ , \quad q={-2 \over 2+3n}\ , \quad \dot u= 
{H' \over H}\partial_r \ , \ee
\be
\sigma_{rr}=-{2 e^{nt} H \over 3}\ ,\quad \sigma_{yy}={e^{(2+n)t}f^2\over 3H}
\ ,\quad \sigma_{zz}={e^{(2+n)t}y^2f^2\over 3H}\ .
\ee
Notice that, depending on the value of  $n$ , the solution contracts and
decelerates or it expands and inflates.

With regard to the dimensionless scalars,
the density parameter $\Omega\equiv 3\mu /{\theta}^2$, the dimensionless
acceleration $W\equiv\dot u/\theta$ and
the shear parameter $\Sigma \equiv 3 \sigma^2 /\theta^2$
 (see \cite{Hewitt88}  for
further details), one has for the first family
\be
\Omega={3(n+1)(n+3) \over (2+3n)n^2(n+2)}{1\over \cosh^2 (\alpha r)}\ 
, \quad W={1\over 2+3n}\tanh (\alpha r) \ ,
\ee
the models then being accelerated
dominated at  large distances; whereas for the second family
\be
\Omega=-{3(n+1)(n+3) \over (2+3n)n^2(n+2)}{1\over \sinh^2 (\alpha r)}\ 
, \quad W={1\over 2+3n}\coth (\alpha r) \ ,
\ee
thus being asymptotically spatially homogeneous. In both
cases 
$ \lim_{r\rightarrow \infty}\Omega = 0 $,
(vacuum dominated models) and   
$\Sigma={1/ (2+3n)^2}$,
 as this quantity is non-vanishing for all possible values of $n$, the models
have no isotropic limit.
 \hfill\break

 \noindent {\bf Case ($B$)}

\begin{center}
\begin{tabular}{|l|c|c|c|c|c|}\hline
$B$ & $k$ & $X$ & $A^2(t,r)$ & $B^2(t,r)$ & $F^2(t,r)$\\\hline
$I$ & $-1,0,1$ & $\partial_r$ & $e^{2r}H^2(t)$ & $e^{2r}H^2(t)$ &
$e^{2r}f^2(t)$ \\ \hline
$II$ & $0$ & $\partial_r-y\partial_y$ & $e^{2nr}H^2(t)$ & $e^{2nr}H^2(t)$ &
$e^{2(n+1)r}f^2(t)$ \\ \hline
\end{tabular}
\end{center}
\begin{center}
{\bf Table 3.3}
\end{center}

\noindent where $r$ in case $BI$ has been re-scaled so as to have $n=1$.

Solving now the field equations for a perfect fluid source, one has:

\[
\begin{array}{|c|c|c|c|c|}\hline
BI & H(t) & f^2(t) & \mu & {\rm Domain} \\\hline\hline
k=\pm 1 & 1 & e^{2t}+{k \over 2} & {\displaystyle k^2 \over\displaystyle 
4e^{2r}f^4} & t \ : \ e^{2t}+{k \over 2}>0 \\\hline
k=0, \pm 1 & 1 & \alpha \sinh 2t +{k \over 2} & {\displaystyle k^2+4 \alpha^2
\over\displaystyle 4e^{2r}f^4} &
t \ : \ \alpha \sinh 2t +{k \over 2}>0 \\\hline
k=\pm 1 & 1 & \alpha \cosh 2t +{k \over 2} & {\displaystyle k^2-4 \alpha^2
\over\displaystyle
 4e^{2r}f^4} &
\begin{array}{c}
\alpha \in (-{1 \over 2}, {1 \over 2}) \\
t \ : \ \alpha \cosh 2t +{k \over 2}>0
\end{array} \\\hline
\end{array}
\]
\begin{center}
{\bf Table 3.4}
\end{center}

\noindent where $\alpha$ is an arbitrary constant, which must be different
from zero  to prevent the occurrence of further Killing vectors.
$k \neq 0$ in both the first and last cases since otherwise one would have a
vacuum
solution in the former case, and negative energy density  in the latter. In
all
three cases, the fluid is irrotational and has an stiff equation of state,
i.e., $\gamma =2$. 
For these solutions one has
 \be
\theta={2\dot f\over e^r f}  \ ,\quad
q={1\over 2}-{3\over 2}{f\ddot f\over (\dot f)^2}\ ,
\quad
\dot u=-\partial_r\ , \quad W={f\over 2\dot f}\ ,
\ee
\be
\sigma_{rr}=-{2 e^r \dot f \over 3f}\ ,\quad \sigma_{yy}={e^rf\dot f\over 3}
\ ,\quad \sigma_{zz}={e^r\Sigma^2 f\dot f\over 3}\ ,
\ee
a dot indicating a derivative with respect to $t$. In all cases the
shear parameter is $\Sigma=1/4$, thus there is no isotropic limit.
For all solutions in $BI$ 
$\lim_{t\rightarrow \infty}W ={1/ 2} $,
thus corresponding to accelerated dominated models, hence with no FRW limit.

\[
\begin{array}{|c|c|c|c|}\hline
BII & H(t) & f^2(t) & \mu  \\\hline\hline
\begin{array}{c} n<-3\\ n>-1
\end{array} & a[S]^{{1\over 2(n+1)}} & S &
{\displaystyle (n+1)(n+3)\over\displaystyle  a^2n^2}e^{-2nr}[S]^{-{2n+3 \over
n+1}} \\\hline -3<n<-1 & a[C]^{{1\over 2(n+1)}}  & C &
-{\displaystyle (n+1)(n+3)\over\displaystyle  a^2n^2}e^{-2nr}[C]^{-{2n+3 \over
n+1}}\\\hline
\end{array}
\]
\begin{center}
{\bf Table 3.5}
\end{center}

\noindent where
\be
a={\rm constant} \ , \quad S=\sinh(2(n+1)t) \ , \quad C=\cosh(2(n+1)t) \ .
\ee
 In both cases the fluid is
irrotational, its equation of state being $p=\mu$, and one has:
\be
\dot u=n\partial_r\ ,
\quad \theta=e^{-nr}H^{-1}\left({\dot H\over H}+2{\dot f\over f} \right)\ ,
\quad \Sigma={n^2\over (2n+3)^2}\ .
\ee
As for the respective dimensionless scalars
\be
\Omega = {3(n+1)(n+3)\over n^2(2n+3)^2}{1\over C^2} \ , \quad
W ={n\over 2n+3}{S\over C}\ ,\ee
\be
q= {-(4n^2+6n+1)\,C^2 +6(n+1)(2n+3)\over
(2n+3)^2 \, C^2} \ ,
\ee
for the first case, and
\be
\Omega = -{3(n+1)(n+3)\over n^2(2n+3)^2}{1\over S^2} \ , \quad
W ={n\over 2n+3}{C\over S}\ , \ee
\be
q= -{(4n^2+6n+1)\,S^2 +6(n+1)(2n+3)\over
(2n+3)^2 \, S^2} \ ,
\ee
for the second case. So, the solutions behave in
different ways depending on the value of the parameter $n$ and the range
 of values of $t$ considered.
\hfill\break

\noindent{\bf Case ($C$)}

\begin{center}
\begin{tabular}{|l|c|c|c|c|c|}\hline
$C$ & $k$ & $X$ & $A^2(t,r)$ & $B^2(t,r)$ & $F^2(t,r)$\\\hline
$I$ & $-1,0,1$ & $\partial_t + \partial_r$ & $e^{t+r}H^2(t-r)$ &
$e^{t+r}L^2(t-r)$ & $e^{t+r}f^2(t-r)$\\\hline
$II$ & $0$ & $\partial_t + \partial_r-y\partial_y$ & $e^{n(t+r)}H^2(t-r)$ &
$e^{n(t+r)}L^2(t-r)$ & $e^{(n+1)(t+r)}f^2(t-r)$\\\hline
\end{tabular}
\end{center}
\begin{center}
{\bf Table 3.6}
\end{center}

\noindent where $t$ and $r$ in $CI$ have been re-scaled so as to make $n=1$.

Unfortunately, no general solutions to the 
field equations for a perfect fluid
can be given in these cases, and as it turns out when trying to solve them
in particular cases, most of the solutions thus found hold only on some
open domains of the manifold of the form $t-r>{\rm constant}$, nevertheless
we next give two solutions that are valid over the whole space-time
manifold, both of them corresponding to the case $CII$:
\hfill\break

\noindent($CII$, $n=-3$)
\be
 f = 1 \ , \quad H^2=\alpha e^{(\beta-4)(t-r)}\ ,  \quad L^2= \alpha
e^{\beta(t-r)}\ .
 \ee
Then one has
\be
\mu=p=e^{3(t+r)}\alpha^{-1}[(4-\beta)e^{(4-\beta)(t-r)}+\beta
e^{-\beta(t-r)}] \ ,
\ee
where $\alpha$ and $\beta$ are constants which can easily be chosen so as to
make $\mu>0$ all over $M$. The dimensionless scalars are
\be
\Omega ={12\over (\beta-7)^2}\left[4-\beta+\beta \, e^{-4(t-r)}\right]
\ , \quad
\theta  ={\beta-7\over 2\sqrt{H}\, e^{-{3\over 2}(t+r)} }\ ,\ee
\be
W ={1-\beta \over \beta-7 }e^{-2(t-r)}\ ,\quad
\Sigma  ={(1-\beta)^2 \over (\beta-7)^2 }\ . 
\ee

\noindent($CII$, $n=1$)
\be
f=1 \ ,\quad H^2=\alpha L^2\ , \quad L^2=\exp \left[
(1-\alpha)(t-r)-{\beta(1-\alpha)\over 2}  e^{-2(t-r)/(1-\alpha)}\right]\ ,
\ee
and then
\bea
(\mu-p)e^{t+r}H^2 &=& 4(1-\alpha)\ ,\\
(\mu+p)e^{t+r}H^2 &=& 2(1-\alpha)\left[ (1-\alpha)+\beta
e^{-2(t-r)/(1-\alpha)}\right]\ ,
\eea
where again $\alpha$ and $\beta$ are constants. The energy conditions
restrict $\alpha$ to values $0<\alpha<1$, and if we demand that the
solution  be valid over the whole manifold, then $\beta$ must be positive.
Notice that in this case, for $\beta \neq 0$, there is no equation of state
of the form $p=p(\mu)$. For $\beta=0$ the solution is an special case of
${\cal H}_5$.
The dimensionless scalars are
\be
\Omega ={3(1-\alpha)\left[ 3-\alpha +\beta\, e^{-{2\over 1-\alpha}(t-r) }
 \right] \over \left[ 3-{\alpha\over 2} +{\beta\over 2} \,
 e^{-{2\over 1-\alpha}(t-r) } \right]^2} \ , \quad
\theta ={  3-{\alpha\over 2} +{\beta\over 2} \,
 e^{-{2\over 1-\alpha}(t-r) }\over \sqrt{H}\, e^{t+r\over 2}}\ , \ee
\be
W = \sqrt{\alpha} \left[ -1 +{6\over 6-\alpha +\beta\,
 e^{-{2\over 1-\alpha}(t-r) } } \right]\ , \quad
\Sigma =\left[ -1 +{6\over 6-\alpha +\beta\,
 e^{-{2\over 1-\alpha}(t-r) } } \right]^2 \ .
\ee
In both cases there is no FRW limit.

As a final remark to this section, notice that expressions appearing in tables
{\bf 3.1}, {\bf 3.3} and {\bf 3.6} are completely general, i.e., valid
regardless the material content.

\section{The $H_3$ case}
In this section we extend a previous work  \cite{Ali1}, correct some errors and
present new solutions.

The existence of a 3-parameter homothetic group $H_3$, implies that of
a $G_2 \subset H_3$ of isometries being the dimensions
of their orbits  three and two respectively (see section 2.1).
 When the Killing subalgebra has null orbits, the metric is of
Kundt's class \cite{Kundt61} and perfect fluids are excluded \cite{Kramer}.  
We shall therefore assume in the sequel that the Killing orbits are non-null.

A classification of all such space-times in terms of the Bianchi type of
the homothetic algebra can be found in \cite{Ali1}.
See also the references cited therein for a (partial) account of papers on
this issue.
 We can summarize the results concerning the topology of the
Killing orbits and the Bianchi type of $H_3$ in the following table:
\[
\begin{array}{|l|c|c|}\hline
G_2-{\rm type} & G_2-{\rm orbits} & H_3-{\rm Bianchi \ type} \\\hline
G_2I & \begin{array}{c} S^1\times \R \\\hline \R^2 \end{array}
& \begin{array}{c} I,II,III \\\hline I,II,III,IV,V,VI,VII \end{array} \\\hline
G_2II & \R^2 & III \\\hline
\end{array}
\]
\begin{center}
{\bf Table 4.1}
\end{center}

As the above table shows, two different topologies are possible in the
Abelian $G_2$ case ($G_2 I$) \cite{Ali1} namely, $V_2$ diffeomorphic to
$\R^2$ or to ${\cal S}^1\times \R$; and it follows in the latter
case that the only possible Bianchi types for ${\cal H}_3$, irrespectively
of the assumed matter content, are $I$, $II$ and $III$ (this holds also if the
${\cal H}_3$ is replaced by a conformal algebra ${\cal C}_3$, see \cite{Mars});
and for the case $V_2\cong\R^2$, the seven soluble Bianchi types can occur.
For the non-Abelian case the only possible homothetic algebra is of the
Bianchi type $III$, and its orbits are diffeomorphic to $\R^2$. 

\subsection{Case $G_2$ Abelian}
In this subsection we shall restrict ourselves to the Abelian case with
spacelike isometric orbits diffeomorphic to $\R^2$, giving  appropriate \lq\lq
translation rules" for the other possibilities. Furthermore, we shall
assume that the Killing orbits admit orthogonal two-surfaces (i.e., orthogonally
transitive $G_2$ metrics).
Cosmological models admitting an Abelian $G_2$ on spacelike orbits have been
studied by Ruiz and Senovilla \cite{Ruiz} and  Van den Berg and Skea
\cite{Bergh92} among others.
 The non-orthogonally transitive case,
Wainwright's classes $A(i)$ and $A(ii)$, have been studied in \cite{Berg91} and
\cite{Wils91} respectively. 

 Adapting two coordinates to two commuting KVs,
say $\xi = \partial_x$ and $\eta = \partial_y$, and choosing two other
coordinates, $t$ and $z$, on the surfaces orthogonal to the isometry
orbits; it follows that the line element can be written in the form (see
for instance \cite{Wainwright2})
\be
ds^2 = -Adt^2 + Bdz^2 + R \left[ F(dx+Wdy)^2 + F^{-1} dy^2 \right]\ ,
\label{5-2}
\ee
where $A$ , $B$ , $R$ , $F$ , and $W$ are all functions of $t$ and $z$
alone.

All the other cases (i.e., timelike Killing orbits and Killing orbits
diffeomorphic to $S^1 \times \R$ of either nature, spacelike or timelike)
can be formally obtained from the above by means of the following
substitutions:

\[
\begin{array}{c|c}
T_2\cong\R^2 & V_2\cong S^1 \times \R \\\hline
\partial_x \mapsto \partial_t & y \mapsto \varphi \\
(t,x)\mapsto i(-x,t) & {\rm Regularity \ condition} \\
W \mapsto iW &  {\rm on \ the  \ axis}
\end{array}
\]
\begin{center}
{\bf Table 4.2}
\end{center}

\noindent where $\varphi$ is the angular coordinate (with the standard
periodicity
$2\pi$).
Regarding solutions with  an Abelian
$G_2$ acting on timelike orbits, (including the astrophysically relevant
stationary and axisymmetric models, which have been studied for many years),
it is worth mentioning  that they 
have  attracted renewed attention, see \cite{Kramer84,Kr90,Mars,Senovilla92}.

It is easy to see from the
commutation relations of the proper HVF, $X$,  with $\xi$ and $\eta$,  and the
homothetic equation specialized to the components $g_{tx}$, $g_{ty}$, $g_{zx}$
and $g_{zy}$ of the metric (\ref{5-2}) that, $X$ must take the form
\bea
X &=& X^t(t,z)\partial_t+X^z(t,z)\partial_z+X^x(x,y)\partial_x+
X^y(x,y)\partial_y  \nonumber\\
&\equiv& \hat X +X^x(x,y)\partial_x+ X^y(x,y)\partial_y \ ,
\eea
where $X^x(x,y)$ and $X^y(x,y)$ are linear functions of their arguments that
yield for every different Bianchi type  the following forms:

\[
\begin{array}{|c|c|c|}
\hline
\ {\rm Type} \ &\  X^x(x,y) \ &\  X^y(x,y) \ \\ \hline
I & 0 &  0 \\
II & y &  0 \\
III & x &  0 \\
IV & x+y &  y \\
V & x &  y \\
VI & x &  qy \\
VII & -y &  x+qy \\
\hline
\end{array}
\]
\begin{center}
{\bf Table 4.3}
\end{center}

As we are interested in perfect fluid solutions for the metric (\ref{5-2}),
it is always possible to perform a change of coordinates in the $t$, $z$ plane
so as to bring the four-velocity of the fluid to a comoving form, preserving
the diagonal form of the  metric  \cite{Wainwright2}. As a
consequence, the fluid flow velocity $u$ can be written as 
\be
u={1\over \sqrt{A}}{\partial \over \partial t} \, \quad {\rm or \
equivalently} \quad u_a=(-\sqrt{A},0,0,0) \ , \label{5-3}
\ee
 the Einstein's field equations taking then a much simpler form. 

In this comoving coordinate chart, and taking into
account the first equation in (\ref{r6}), it is easy to see that the part of
the homothetic vector field
 orthogonal to the Killing orbits, $\hat X$, is
\be
\hat X = X^t(t)\partial_t +X^z(z)\partial_z \ .
\ee
We can now use the remaining coordinate freedom in the $t$, $z$ plane
$[t\rightarrow m(t)\, , \ z\rightarrow n(z)]$ to bring $\hat X$ to
either of the following three forms
\bea
&(i)&\hat X=\partial_t \ ,\label{5-6}\\
&(ii)&\hat X=\partial_r \ ,\label{5-7}\\
&(iii)&\hat X=\partial_t+\partial_r \ .\label{5-8} \eea

Thus, three classes of perfect fluid solutions arise, depending on the
orientation of the fluid flow $u$ relative to the homothetic orbits:
\begin{quotation}
$(i)$ The fluid flow is tangent to the homothetic orbits, and they are
then timelike.

$(ii)$ The fluid flow is orthogonal to the homothetic orbits, and therefore
they are spacelike.

$(iii)$ \lq\lq Tilted" fluid flow, i.e., $u$ is neither tangential to nor
orthogonal to the homothetic orbits, which are then not constrained -a
priori- to being timelike or spacelike, and so that  their nature may vary from
point to point over the space-time.
\end{quotation}

\subsubsection{Fluid flow tangent to the homothetic orbits}
This case, assuming the existence of two hypersurface orthogonal KVs (i.e.,
diagonal metric)  has been   thoroughly studied by
Wainwright, Hewitt, and collaborators in an interesting series of articles
\cite{Hewitt88,Hewitt91,Hewitt90}, where the properties of these models are
analyzed using the qualitative theory of plane autonomous systems,
 showing
that (first-class) self-similar solutions within this family can represent
the asymptotic states at later times of more general  inhomogeneous $G_2$
models.
  Uggla \cite{Uggla92} found four explicit solutions of this type.
We shall not give here any explicit solution belonging to this class (see
the above references and those cited therein), but rather provide the general
form  of the metric functions (including the non-diagonal cases)  and
briefly discuss the generic behaviour of the kinematical quantities associated
with the fluid (acceleration, deceleration parameter, shear,...).

The metric (in comoving coordinates) for the case of four-velocity
tangent to the $H_3$ orbits (i.e., form ($i$) of the proper homothetic
vector field) takes the
form: 
\be ds^2=e^{2nt}\{-A(z)dt^2+B(z)dz^2 +e^{\alpha t}R(z) [F(t,z)(dx+
W(t,z)dy)^2 +F^{-1}(t,z)dy^2] \}
\label{5-9}
\ee
where $n$ is the homothetic constant. By rescaling the coordinate $z$ one
can set
\be
A(z)=B(z) \ .
\ee
Then, the functional form of the metric functions  can 
be worked out for each Bianchi type. Thus, one has:
\bea
(I) & & \alpha =0\ , \quad F=f(z)\ , \quad W=w(z) \ , \label{5+9}\\
(II) & & \alpha =0\ , \quad F=f(z)\ , \quad W=w(z)-t \ , \\
(III) & & \alpha =-1\ , \quad F=e^{-t}f(z)\ , \quad W=e^t w(z) \ , \\
(IV) & & \alpha =-2\ , \quad F=f(z)\ , \quad W=w(z)-t \ , \\
(V) & & \alpha =-2\ , \quad F=f(z)\ , \quad W=w(z) \ , \\
(VI) & & \alpha =-(1+q)\ , \quad F=e^{-(1-q)t}f(z)\ , \quad
W=e^{(1-q)t}w(z) \ , \\ 
(VII)& & \alpha=-q \ ,\label{5++9}\eea
$$F={2 \over \sqrt{4-q^2}} \left[\sqrt{1+c(z)^2+g(z)^2}
+c(z)\cos(\sqrt{4-q^2}t) +g(z)\sin(\sqrt{4-q^2}t) \right]\ ,$$
 $$W={q\over 2}+{{\sqrt{4-q^2}\over
2}[-g(z)\cos(\sqrt{4-q^2}t)+ c(z)\sin(\sqrt{4-q^2}t)] \over
\sqrt{1+c(z)^2+g(z)^2} +c(z)\cos(\sqrt{4-q^2}t) +g(z)\sin(\sqrt{4-q^2}t)}\ .
$$
Notice that the form of these functions again holds for any energy-momentum
tensor,  since no use has been made of
the field equations in deducing them.

The cases studied by Wainwright and collaborators correspond to types $I$,
$III$, $V$, and $VI$, since these are the only ones in which the function $W$
can be set equal 
to zero. For type $VII$, $W={q\over 2}$ implies the existence
of a further Killing vector
 tangent to the Killing orbits and the metric would then admit
a multiply transitive group $H_4$ of homotheties.
Notice also, that the diagonal cases are  separable in the variables $t$,
$z$,  thus being
special cases of the solutions studied by Ruiz and Senovilla \cite{Ruiz} or
those of Agnew and Goode \cite{Agnew} for $\gamma=2$.

Specializing equation (\ref{r6}) to the matter variables $\mu$ and $p$, we
obtain \be
\mu=e^{-2nt}\hat \mu(z) \ , \quad p=e^{-2nt}\hat p(z) \ , \label{5-10}
\ee
and by computing the kinematical quantities associated to 
the fluid velocity vector
(\ref{5-3}) for the metric (\ref{5-9}) 
\be
\theta={3n+\alpha\over e^{nt}\sqrt{A}} \ ,\quad  {\bf q}={-\alpha\over
3n+\alpha} \ , \quad\dot u={1\over 2}{A'\over A}dz \ , \label{5-11}
\ee
where a dash denotes a derivative with respect to $z$ and ${\bf q}$ here
denotes the deceleration parameter. The non-vanishing components of the shear
tensor are \be
\sigma_{zz}= -e^{nt}\sqrt{A} {\alpha \over 3}\ , \quad
\sigma_{xx}=e^{nt}{RF\over 6\sqrt{A}}\left(3{\dot F \over F} +\alpha\right)\ ,
\ee
\be
\sigma_{xy}=e^{nt}{RF\over 6\sqrt{A}}\left(3W{\dot F \over F}
+W\alpha +3\dot W\right)\ , 
\ee
\be
\sigma_{yy}={e^{nt}\over 6\sqrt{A}}{R\over F}\left(-3{\dot F \over F} +\alpha
+3W^2F\dot F+ W^2F^2\alpha + 6F^2W\dot W \right)\ , 
\ee
thus,  the shear scalar is
\be
\sigma^2={3 \left({\dot F\over F} \right)^2 +\alpha^2 +3F^2\dot W^2
\over 12 A e^{2nt}}\ .
\ee

If one assumes that the fluid has an equation of state of the form (\ref{r7}),
from the contracted Bianchi identities  it follows
\be
\gamma(\dot \mu+\mu \theta)u^a + \gamma\mu\dot u^a + 
(\gamma-1)\mu_{,b}g^{ba}=0 \ , 
\ee
where $\dot \mu\equiv \mu_{,t}u^t$. Contracting the above expression with
$u^a$, one gets
\be
\dot \mu= -\gamma\mu\theta \ . \label{5-12}
\ee
Assuming $\theta\not= 0$ (the case $\theta=0$, although mathematically
possible, is not physically interesting since it would correspond to a
non-expanding universe, thus contradicting observations) and substituting
(\ref{5-10}) and (\ref{5-11}) in equation (\ref{5-12}), it follows
\be
\gamma={2n\over 3n+\alpha}\ .\label{5-13}
\ee
Specializing the quantities $\gamma$ and ${\bf q}$ to each Bianchi type, one
gets

\begin{center}
\begin{tabular}{|c|c|c|c|c|c|c|c|}\hline
 Type & $I$ & $II$ & $III$ & $IV$
& $V$ & $VI$ & $VII$ \\ \hline\hline
$\gamma$ &${\displaystyle 2\over \displaystyle 3}$ &${\displaystyle 2\over
\displaystyle 3}$ &${\displaystyle 2n\over\displaystyle 3n-1}$ &
${\displaystyle 2n\over\displaystyle 3n-2}$ &${\displaystyle 2n\over
\displaystyle 3n-2}$ &${\displaystyle 2n\over\displaystyle 3n-(1+q)}$ &$
{\displaystyle 2n\over \displaystyle 3n-q}$ \\ \hline ${\bf q}$ & $0$ &
$0$ & ${\displaystyle 1\over\displaystyle 3n-1}$ & ${\displaystyle 2\over
\displaystyle 3n-2}$ & ${\displaystyle 2\over \displaystyle 3n-2}$ &
${\displaystyle 1+q\over\displaystyle 3n-(1+q)}$ & ${\displaystyle q\over
 \displaystyle 3n-q}$ \\ \hline
\end{tabular}
\end{center}
\begin{center}
{\bf Table 4.4}
\end{center}

Notice that the only shear-free solution is the type $I$ one. For types $II$ to
$VI$ there is no limit where $\Sigma\equiv 3\sigma^2/\theta^2$ becomes zero;
and for type $VII$, $\Sigma\rightarrow 0$ for some spatial limit if and only
if $c(z)\rightarrow 0$, $g(z)\rightarrow 0$ and $q=0$ (i.e., $F\rightarrow 1$
and $W\rightarrow 0$). Thus, only types $I$ and $VII$ can have solutions
with a FRW limit, but in both cases $\mu+3p=0$ and ${\bf q}=0$, so they
are not very relevant from a physical point of view.

\subsubsection{Fluid flow orthogonal to the homothetic orbits}

The case of spatially homothetic orbits was thoroughly studied by Eardley
\cite{Eardley} where a classification scheme of these models was given and
their dynamical properties were studied. Luminet \cite{Luminet} constructed
a convenient basis of 1-forms and gave its explicit form in terms of a
standard coordinate basis $\{dx^a\}$ as well as  the expression of the
homothetic vector in the dual basis $\{\partial /\partial x^a\}$. He also
proved a theorem showing that perfect fluid models of a certain class were
incomplete in the sense of Hawking and Ellis \cite{Hawking73}.

For the sake of completeness, we just mention that the form of the
metric, corresponding to a homothetic vector field
 of the form ($ii$), can be formally obtained
from (\ref{5-9}) to (\ref{5++9}) by simply reversing the roles of
the coordinates $t$ and $z$.

The expressions of the acceleration, expansion, deceleration parameter and
shear scalar are given by
\be
\dot u=ndz \ , \quad \theta={1\over 2 e^{nz}\sqrt{A}}\left\{{\dot A\over A}+
2{\dot R\over R} \right\} \ ,
\ee
\be
q={2 \left[4\left( {\dot A\over A} \right)^2 +{\dot A\dot R\over AR} +
4\left( {\dot R\over R} \right)^2 - 3{\ddot A\over A} -6{\ddot R\over R}
\right] \over \left( {\dot A\over A}+ 2{\dot R\over R} \right)^2} \ ,
\ee
\be
\sigma^2={\left( {\dot A\over A}-{\dot R\over R} \right)^2 + 3 \left( {\dot
F\over F} \right)^2+ 3F^2(\dot W)^2\over 12 A e^{2nz}} \ .
\ee

In this case, the homothetic vector field,
$X$, and the four-velocity $u$ are mutually orthogonal,
thus if a barotropic equation of state is assumed, then necessarily $p=\mu$,
i.e., stiff-matter and by simple inspection of the field equations this is
seen to be equivalent to:
 \be
{\ddot R\over R}=(2n+\alpha)^2 \ .
\ee

\subsubsection{\lq\lq Tilted" fluid flow}

Finally, the form ($iii$) for the proper homothetic vector field
 is precisely the case we are
currently interested in, namely, $u$ not tangent nor orthogonal to the
homothetic orbits.

Specializing now the homothetic equation to the metric (\ref{5-2}) we obtain
\bea
ds^2&=& e^{n(t+z)}\left\{ -A(t-z)dt^2+B(t-z)dz^2\right. \nonumber\\
& &\left. +e^{\alpha
{t+z\over 2}} R(t-z) \left[F(dx+Wdy)^2+F^{-1}dy^2 \right]\right\} \ ,
\eea
where again $n$ is the homothetic constant. The parameter $\alpha$ and the
functional form of the metric functions for each Bianchi type, are those
given by (\ref{5+9})-(\ref{5++9}) after effecting the following
substitutions
\be
t\longmapsto {t+z\over 2}\ , \qquad z\longmapsto t-z\ .
\ee

The kinematical quantities being
\be
\dot u={1\over 2}\left(n-{A'\over A} \right)dz \ , \quad
\theta={3n+\alpha +{B'\over B}+2{R'\over R}\over 2 e^{n(t+z)/2}\sqrt{A}} \
,
\ee
\be
\sigma^2={\left({\alpha \over 2}-{B'\over B}+{R'\over R}  \right)^2 +3
\left({F_{,t}\over F} \right)^2+ 3F^2(W_{,t})^2 \over 12 e^{n(t+z)}A} \ ,
\ee
where a dash denotes here a derivative with respect to $(t-z)$.  A careful 
study shows
that there are no shear-free solutions in this case admitting a maximal
group $H_3$ of homotheties: shear-free solutions are not possible for types
$II$ and $IV$; for the Bianchi types $I$, $V$ and $VII$, the shear-free
condition
 implies that the
functions $F$ and $W$ must be constants and therefore a further 
Killing vector tangent to
the Killing orbits occurs; for types $III$ and $VI$, $W$  must vanish, the
type $III$  solution then being a homogeneous Bianchi $VI$ model, and the type
$VI$ one such that $\mu+p=0$, thus  not corresponding to a perfect fluid.

Notice that in \cite{Ali1}, the homothetic constant $n$ was set
equal to $1$ from the beginning, before choosing the coordinates in the
surfaces orthogonal to the Killing orbits. By doing so, some  solutions
were left out, since, although one can always re-scale the proper HVF $X$
with a factor ${1 / n}$, such an scaling can not always be reabsorbed
by a redefinition of the coordinates. We correct here that error.

As was pointed out before,
all diagonal ($W=0$), perfect fluid
solutions  (admitting an orthogonally transitive
Abelian $G_2$ with flat spacelike orbits) and such that the metric functions
$A$, $B$, $R$ and $F$ are separable in the variables $t$ and $z$ are already
known \cite{Ruiz,Agnew}.

We shall next present some new exact solutions not included in \cite{Ali1},
which have been obtained  assuming $W=0$ (diagonal), but which are not
of separable variables in the above sense.
\hfill\break

Type $III$:
\be
 R=1\ ,\quad f=e^{\lambda (t-z)}\ , \label{47}
\ee
$$
{A\over B}={1-2\lambda \over 1+2\lambda}\ ,  \quad A=a \exp\left( n(t-z)+
{\alpha \over {1 \over 2} -n}e^{({1 \over 2}-n)(t-z)} \right)\ ,
$$
where $0<\lambda<1/2$ and $\lambda^2+n^2=1/2$ and \bea
(\mu-p)Ae^{n(t+z)}&=&{4\lambda \over 1+2\lambda}\left(  {1\over
2}-n\right)^2\ ,\nonumber
\\ (\mu+p)Ae^{n(t+z)}&=&{4\lambda \over 1+2\lambda}\left( n- {1\over 2}
\right) {A'\over A}\ ,\nonumber
\eea
 $\lambda$,
$\alpha$ and $a$ are constants. In order to satisfy the energy-conditions,
if $n>0$, $\alpha$ must also be positive and the value of $n$ is then
restricted to $(1/2\, , \, \sqrt{2}/2 )$; for negative values of $n$,
$\alpha$ is also negative and $n\in(-\sqrt{2}/2, -1/2)$, in these
cases  a barotropic equation of state does not exist. When $\alpha=0$
the solution is a particular case of a homogeneous Bianchi $VI$
model. 
\hfill\break

Type $III$:
\be
 R=1\ ,\quad f=e^{\lambda (t-z)}\ , \label{48}
\ee
$$
{A\over B}=e^{(2n-1)(t-z)}\ , \quad \lambda^2=(n-1)^2\  , \quad
 n\not={1\over 2}\ , 
$$
$$
B=a \exp\left(-{\lambda+ 2(n-1)^2\over 2n-1}(t-z) \right)
\left[ \exp\left(-(2n-1)(t-z)\right)-1 \right]^c\ ,
$$
where $a$ and $c$ are constants
$$
\mu=p ={1 \over 2A}\left[2n-{3+c\over 2}\right]e^{-n(t+z)}\ .
$$
\hfill\break

Type $III$:
\be
 R=e^{\lambda(t-z)}\ , \quad f=e^{-{1 \over 2}(t-z)}\ , \quad
A=a^2e^{(2\lambda+2n-1)(t-r)}\ , \quad B=b^2\ , \label{49}
\ee
where $a$ and $b$ are constants, $\lambda=-(2n-1)+\sqrt{3n^2-2n+1/4}$,
and
\bea
(\mu-p)e^{n(t+z)}&=&{1\over B}\lambda(2n-1-2\lambda)\ , \nonumber\\
(\mu+p)e^{n(t+z)}&=&{1\over A}(n^2+{\lambda\over 2}-{3\over 4})+{1\over
B}\lambda(\lambda+2n-{3\over 2})\ .\nonumber
\eea
  {}From
 where it follows that, in order that the solution satisfies the weak
and dominant energy conditions all over the manifold, one must have
 $$n\in(0.8210368162407501..., {3\over 2})\ ,$$ and, again, a barotropic
equation of state $p=p(\mu)$ does not exist,  except in the case $n=3/2$,
when $\mu=p$. 
\hfill\break

Type $V$:
\be
ds^2={e^{t+z}\over f_o^2}\vert \varphi\vert^{2c^2+2c}\left\{
{-M^4\varphi^2 dt^2+dz^2\over M^4\varphi^2-1}\right\}+\vert
\varphi\vert^{-2c}dx^2+\vert \varphi\vert^{2c+2}dy^2 \ ,  \label{51}
\ee
$$
\mu=p={f_o^2\over e^{t+z}}{M^4\varphi^2-1 \over 2M^2 \vert
\varphi\vert^{2c^2+2c+2}} \ , \quad n=1 \ , $$
where $c$, $M$ and $f_o$ are constants,  and $\varphi$ is a function of
$t-z$  given implicitly by 
$$M^2(t-z)=\ln \vert\varphi\vert-{M^4\over
2}\varphi^2 \ . $$
\hfill\break

Type $V$:
\be
 R=1\ , \quad f=e^{\lambda (t-z)}\ , \label{52}
\ee
$$
{A\over B}=e^{2(n-1)(t-z)}\ , \qquad \lambda^2=(n-1)(n-3)\ ,$$
$$
A=a e^{(n+1)(t-z)}\left[1-e^{-2(n-1)(t-z)}\right]^{c\over 2(n-1)}\ ,$$
$$
p=\mu ={(n-1)(4-c) \over 2A} e^{-n(t+z)}\ ,$$
with $a$ and $c$ constants, and $n$ restricted to being $n<1$ or $n>3$ in order
to satisfy the energy conditions.
\hfill\break

For type $VI$ two solutions have been obtained. For the sake of simplicity,
we will give them in non-comoving coordinates. Thus
\be
ds^2={e^{2t}\over F^2}\left\{-dt^2+dz^2 +e^{-2t}B^2 dx^2
+e^{-2qt}S^2 dy^2 \right\}\ ,
\ee
\be
u_t=-{e^t\over F} \cosh a \ , \quad u_z={e^t\over F}\sinh a \ .
\ee

The first solution is
\be
F=f_oB\ , \quad S=C^{1-{q\sin c\over 1-q}}E \ , \quad B=C^{-{q\sin c\over
1-q}}E \ ,  \label{53}\ee
$$\mu=p=e^{-2t}2q(1-q)f_o^2\alpha C^{-2-2{q\sin c\over 1-q}}(A\cos c
+2\alpha \sin c)E^2 \ , $$
where
$$A=\alpha^2 e^{(1-q)z}-e^{-(1-q)z} \ , \quad C=\alpha^2
e^{(1-q)z}+e^{-(1-q)z} \ ,$$
$$
E=\exp\left[{2q\cos c\over 1-q}\tan^{-1}\left(\alpha e^{(1-q)z} \right)
\right] Ê\ ,$$
and
$$\cosh a={\sqrt{1+\sin c}\, C \over \sqrt{4\alpha A\cos c+8\alpha^2\sin
c}} \ , \quad \sinh a={-(1+\sin c)A+2\alpha\cos c\over \sqrt{1+\sin c} 
\sqrt{4\alpha A\cos c+8\alpha^2\sin c}} \ , $$
where $c$, $\alpha$ and $f_o$ are constants. In order to have positive
energy density, the parameter $q$ is restricted to $q\in(0,1)$. A
particular, simpler case can be obtained by choosing $\cos c=0$ and $\sin
c=1$.

The other solution is
\be
F=f_oB\ , \quad S=R^{1+{q^2\over c(1-q)}}T^{1+{c\over 1-q}} \ , \quad
B=R^{{q^2\over c(1-q)}}T^{{c\over 1-q}} \ , \label{54}\ee
$$
\mu=p=e^{-2t}{q-1\over c}\alpha^2f_o^2 \left[
c^2R^2-q^2T^2\right] R^{2{q^2\over c(1-q)}-2}T^{{2c\over 1-q}-2} \ ,  $$
where
$$R=e^{{1-q\over 2}z}-\alpha^2 e^{-{1-q\over 2}z}\ , \quad T=e^{{1-q\over
2}z}+\alpha^2 e^{-{1-q\over 2}z}\ ,$$
and
$$\cosh a={(c-q)TR\over 2\alpha \sqrt{c^2R^2-q^2T^2}}\ , \quad \sinh
a={qT^2-cR^2\over  2\alpha \sqrt{c^2R^2-q^2T^2}}$$
again  $c$, $\alpha$ and $f_o$ are constants. Notice that the solution is
only valid for 
$$ c^2R^2-q^2T^2>0\ .$$

Notice that  the solutions that have a stiff matter equation of state
($p=\mu$)  can be derived from vacuum solutions (also admitting an
Abelian $G_2$) using a method proposed by Wainwright et al. \cite{Wainwright79}

\subsection{Non-Abelian $G_2$}

For a non-Abelian $G_2$, a local system of coordinates can be chosen in which
the  Killing vectors,
 say $\xi$ and $\eta$, are
\be
\xi=\partial_1 \ , \quad
\eta= x^1\partial_1+\partial_2 \ . \label{n-2}
\ee
Now, supposing the existence of a proper homothetic vector field,
 say $X$, one can see that the
only allowed Bianchi type for $H_3$ is  $III$, i.e.:
\be
[\xi,\eta]=\xi \ , \quad [\xi,X]=0 \ , \quad [\eta, X]=0 \ . \label{n-3}
\ee
Taking into account (\ref{n-2}) and (\ref{n-3}) one easily comes to the 
following form
of $X$:
\be
X^a=(e^{x^2}X^1(x^3,x^4),X^2(x^3,x^4),X^3(x^3,x^4),X^4(x^3,x^4))\ .
\label{n-4} \ee
Note that, since the orbits associated with $H_3$ and $G_2$ can not coincide
\cite{Hall90}, the components $X^3$ and $X^4$ of the homothetic vector 
field 
cannot both vanish.
Also, notice that the homothetic constant can in this case be set equal 
to one without altering our choice of coordinates.

\subsubsection{The orthogonally transitive case}
As in the previous Abelian case, we will restrict our attention just to the
orthogonally transitive $G_2$ metrics. Regarding non-orthogonally transitive
 $G_2$, a
discussion can be found in \cite{Bergh88} where a study of perfect fluid
solutions with four-velocity orthogonal to the isometric orbits is given. In
that reference it is also assumed that the Killing vector
$\xi$ is hypersurface orthogonal
 and the homothetic vector field,
 $X$, is orthogonal to the fluid velocity.
These assumptions imply that the fluid is to be \lq\lq stiff" ($p=\mu$) without
any a priori assumption of an equation of state. No explicit solutions 
are known so far,
 but it is shown that solutions with pressure and matter positive on
an open set can in principle exist
 by suitably specifying the initial
conditions.

For orthogonally transitive $G_2$, the metric can be written as
\be
 g_{ab}= \left(\begin{array}{cccc}
e^{-2x^2} a_{11} & e^{-x^2} a_{12} & 0 & 0 \\
e^{-x^2} a_{12} &  a_{22} & 0 & 0 \\
0 &  0 & a_{33} & 0 \\
0 & 0  &  0  & \epsilon
\end{array}
\right) 
\label{n-5}
\ee
where $\epsilon = \pm 1$ and $a_{ij}=a_{ij}(x^3,x^4)$.

For this case, on can see that, assuming non-null homothetic orbits $V_3$, it
is always possible to perform a coordinate change in the two-spaces orthogonal
to the Killing orbits, such that it brings the homothetic vector field
 to the form
\be
X^a=(e^{x^2}X^1(x^3,x^4),X^2(x^3,x^4),1,0)\ .
\label{n-6} \ee
and the line element can be written as
\be
ds^2=A(e^{-x^2}\, dx^1+W\, dx^2)^2 +B(dx^2)^2 +
F \left( (dx^3)^2 +\epsilon (dx^4)^2 \right)\ , \label{n-7}
\ee
where  $A$, $B$, $F$, and $W$ are functions of $x^3$
and $x^4$ alone.

Specializing now
 the homothetic equation to the homothetic vector (\ref{n-6}) and the metric
(\ref{n-7}) yields  the following forms for $X$ and the metric functions
\be
X^a=(\alpha e^{x^2}, n,1,0)\ , \quad \alpha, n={\rm const} \ , \label{n-8}
\ee
\be
A=e^{2(1+n)x^3}a(x^4)\ , \quad B=e^{2x^3}b(x^4) \ , \quad F=e^{2x^3}f(x^4) \ , 
\label{n-9}\ee
and
\be
W=\left\{
\begin{array}{ll}
-{\displaystyle\alpha \over\displaystyle n}+e^{-nx^3}w(x^4)\ & n\not= 0\\
-\alpha x^3+w(x^4)& n=0
\end{array}  \right. \label{n-10}
\ee
In the case  $n\not=0$, one can still perform the coordinate change
\be
\hat x^1= x^1-{\alpha \over n}e^{x^2} \ , \label{n-11}
\ee
so that the homothetic vector field and $W$ take the forms
\be
X^a=(0,n,1,0)\ , \quad
W=e^{-nx^3}w(x^4)\ . \label{n-13}
\ee
It can be easily shown, just by computing the Einstein and Riemann tensors, that
the only vacuum solution with a non-Abelian group of isometries acting
orthogonally transitively and admitting a proper homothetic vector field,
is Minkowski space-time.

For perfect fluid solutions, equation (\ref{r6}) specialized to the Killing
vectors
 (\ref{n-2}) implies for the fluid velocity 
 \be
u_a=\left(e^{-x^2}u_1(x^3,x^4),u_2(x^3,x^4),u_3(x^3,x^4),u_4(x^3,x^4)\right)\ .
\label{n-14} \ee
For orthogonally transitive $G_2$ models, it is possible to perform a change of
coordinates in the $x^3$, $x^4$ plane so as to write the four velocity and the
metric as 
\be
u_a=\left(e^{-x^2}\tilde
u_1(x^3,x^4),\tilde u_2(x^3,x^4),0,\tilde u_4(x^3,x^4)\right)\ ,
\label{n-15} \ee
\bea
ds^2&=&A(x^3,x^4)(e^{-x^2}\, dx^1+W(x^3,x^4)\, dx^2)^2 +B(x^3,x^4)(dx^2)^2 
\nonumber \\
& &+F(x^3,x^4)(dx^3)^2 +G(x^3,x^4)(dx^4)^2 \ , \label{n-16}
\eea
and, as a consequence, the field equations take  on a much simpler
form. In this coordinate chart and taking into account (\ref{r6}) and 
(\ref{r1}) specified to the components $g_{13}$, $g_{14}$, $g_{23}$ and
$g_{24}$, it is easy to see that $X$ must be of the form
\be
X^a=(\alpha e^{x^2}, n,X^3(x^3), X^4(x^4)) \ , \label{n-17}
\ee
and the following possibilities then arise
\bea
&{\rm I}& \quad X^a=(\alpha e^{x^2}, n,0, 1) \ , \label{n-18}\\
&{\rm II}& \quad X^a=(\alpha e^{x^2}, n,1,0) \ , \label{n-19}\\
&{\rm III}& \quad X^a=(\alpha e^{x^2}, n,1, 1) \ , \label{n-20}
\eea
In the three cases, for $n\not=0$ one can perform the coordinate change
(\ref{n-11}) thus enabling one to set $\alpha$ zero.

The homothetic equation specialized to the metric (\ref{n-16}) yields then the
following possibilities
\[
\begin{array}{|l|c|c|c|}\hline
{\rm type} & X^a & A(x^3,x^4) & W(x^3,x^4) \\ \hline\hline
{\rm I}.a& (0,n,0,1)& e^{2(1+n)x^4}a(x^3) & e^{-nx^4}w(x^3) \\ \hline
{\rm I}.b& (\alpha e^{x^2}Ê,0,0,1)& e^{2x^4}a(x^3) & -\alpha x^4+w(x^3) \\ \hline
{\rm II}.a& (0,n,1,0)& e^{2(1+n)x^3}a(x^4) & e^{-nx^3}w(x^4) \\ \hline
{\rm II}.b& (\alpha e^{x^2}Ê,0,1,0)& e^{2x^3}a(x^4) & -\alpha x^3+w(x^4) \\
\hline {\rm III}.a& (0,n,1,1)& e^{(1+n)(x^3+x^4)}a(x^3-x^4) &  e^{-{n\over
2}(x^3+x^4)}w(x^3-x^4) \\ \hline 
{\rm III}.b & (\alpha e^{x^2}Ê,0,1,1) & e^{x^3+x^4}a(x^3-x^4)
& -{\alpha\over 2}(x^3+x^4)+ w(x^3-x^4) \\ \hline
\end{array}
\]
\begin{center}
{\bf Table 4.5}
\end{center}

\[
\begin{array}{|l|c|c|c|}\hline
{\rm type} & B(x^3,x^4)  & F(x^3,x^4) & G(x^3,x^4) \\ \hline\hline
{\rm I}&   e^{2x^4}b(x^3) &  e^{2x^4}f(x^3)& \epsilon e^{2x^4}f(x^3)\\ \hline
{\rm II}&  e^{2x^3}b(x^4) &  e^{2x^3}f(x^4)& \epsilon e^{2x^3}f(x^4)\\ \hline
{\rm III}&  e^{x^3+x^4}b(x^3-x^4) &  e^{x^3+x^4}f(x^3-x^4)& 
e^{x^3+x^4}g(x^3-x^4) \\ \hline
\end{array}
\]
\begin{center}
{\bf Table 4.6}
\end{center}
and for the velocity field one has
\[
\begin{array}{|l|c|c|c|}\hline
{\rm type} &\tilde u_1(x^3,x^4) &\tilde u_2(x^3,x^4) & \tilde u_4(x^3,x^4) \\
\hline\hline
{\rm I}.a& e^{(1+n)x^4}\hat u_1(x^3) & e^{x^4}\hat u_2(x^3)& e^{x^4}\hat u_4(x^3)\\
\hline 
{\rm I}.b& e^{x^4}\hat u_1(x^3) &-\alpha x^4e^{x^4}\hat u_1(x^3)+e^{x^4}\hat u_2(x^3)
& e^{x^4}\hat u_4(x^3)\\ \hline 
{\rm II}.a& e^{(1+n)x^3}\hat u_1(x^4) & e^{x^3}\hat u_2(x^4)& e^{x^3}\hat
u_4(x^4)\\ \hline 
{\rm II}.b& e^{x^3}\hat u_1(x^4) &-\alpha x^3e^{x^3}\hat u_1(x^4)+e^{x^3}\hat
u_2(x^4) & e^{x^3}\hat u_4(x^4)\\ \hline 
{\rm III}.a& e^{{1+n\over 2}(x^3+x^4)}\hat u_1(v) & e^{{x^3+x^4\over 2}}\hat
u_2(v)& e^{{x^3+x^4\over 2}}\hat u_4(v)\\ \hline 
{\rm III}.b& e^{{x^3+x^4\over 2}}\hat u_1(v) & 
-\alpha {x^3+x^4\over 2}e^{{x^3+x^4\over 2}}\hat u_1(v) +
e^{{x^3+x^4\over 2}}\hat u_2(v)& e^{{x^3+x^4\over 2}}\hat u_4(v)
\\ \hline 
\end{array}
\]
\begin{center}
{\bf Table 4.7}
\end{center}
where the subcases $a$ and $b$ refer to whether $n\not= 0$ or $n=0$
respectively, and $v$ stands for $x^3-x^4$.

Since $u_3=0$ in this coordinate chart, 
 the components of the Einstein tensor $G_{13}$, $G_{23}$ and $G_{34}$
must vanish identically, hence
\be
W_{,3}=0\ , \label{n-24}
\ee
\be
{A_{,3}\over A}={B_{,3}\over B}\ , \label{n-25}
\ee
\be
0={B_{,3}\over B}\left(-{1\over 4}{A_{,4}\over A}+ {1\over 2}{F_{,4}\over F}
+{3\over 4}{B_{,4}\over B}\right) +{1\over 4}{G_{,3}\over G} \left( 
{A_{,4}\over A}+{B_{,4}\over B} \right) -{B_{,34}\over B}\ . \label{n-26}\ee
 where $_{,i}$ means derivative with respect $x^i$.
  {}From
 these equations we found more explicit forms for the metric functions,
that are given by: \hfill\break

{\bf Case ${\bf I.a}$}
\bea
ds^2&=& e^{2x^4}\left\{ k\, e^{2nx^4} \left[ f(x^3)\right]^{2+n\over n}
\left(e^{-x^2}dx^1+w\, e^{-nx^4}dx^2 \right)^2 +b \left[
f(x^3)\right]^{2+n\over n} \left(dx^2\right)^2 \right.\nonumber \\
& & \left. +f(x^3)\left( \left(dx^3\right)^2 +\epsilon \left(dx^4\right)^2 
\right) \right\}\ , \label{n-27}
\eea
where $k=\pm 1$, $w$ and $b$ are arbitrary constants. Note that $n$ must be
different from zero or a third Killing vector
 occurs and the metric becomes then LRS. In
such a case the metric functions $a(x^3)$ and $b(x^3)$ are still
proportional to each other, but no relation exits, in principle, with $f(x^3)$.
\hfill\break

{\bf Case ${\bf I.b}$}
\be
ds^2=  e^{2x^4}\left\{  a(x^3)\left(e^{-x^2}dx^1-\alpha x^4 dx^2 \right)^2
+b\,  a(x^3) \left(dx^2\right)^2 + \left(dx^3\right)^2+ \epsilon
\left(dx^4\right)^2 \right\}\ , \label{n-28}\ee
where $b$ is an arbitrary constant different from zero to prevent a
non-singular metric, and $\alpha\not=0$ if the group $G_2$ is to be maximal.
\hfill\break

{\bf Case ${\bf II}$} \hfill\break
 Only one possibility arises in this case, namely  $\alpha =n=0$; thus the
line element  becomes
\be
ds^2=  e^{2x^3}\left\{  a(x^4)\left(e^{-x^2}dx^1+ w(x^4)\, dx^2 \right)^2
+b(x^4) \left(dx^2\right)^2 + \left(dx^3\right)^2+ \epsilon
\left(dx^4\right)^2 \right\}\ , \label{n-29}\ee
\hfill\break

{\bf Case ${\bf III.a}$}
\bea
ds^2&=& e^{x^3+x^4}\left\{ e^{n(x^3+x^4)}a(v)  \left(e^{-x^2}dx^1+w\,
e^{-nx^4}dx^2 \right)^2 + \right. \nonumber \\
& &\left. b e^{n(x^3-x^4)}a(v) \left(dx^2\right)^2 + f(v)
\left(dx^3\right)^2 +g(v) \left(dx^4\right)^2 \right\}\ , \label{n-30}
\eea
$w$ being an arbitrary constant and  $v=x^3-x^4$. 
\hfill\break

{\bf Case ${\bf III.b}$}
\be
ds^2= e^{x^3+x^4} \left\{ a(v) \left[ \left(e^{-x^2}dx^1-\alpha x^4 dx^2
\right)^2 +b\left(dx^2\right)^2 \right] + f(v)
\left(dx^3\right)^2 +g(v) \left(dx^4\right)^2 \right\}\ . \label{n-31}
\ee
Again, $\alpha$ and $n$ are arbitrary non-null constants. The differential
equation (\ref{n-26}) for case $III$ can be rewritten as
\be
0={g'\over g}\left(1-{a'\over a} \right)- {f'\over f}\left(1+n+{a'\over a}
\right) +1+n{a'\over a}+ \left({a'\over a} \right)^2 +2 \left({a'\over a}
\right)'\ , \label{n-32}
\ee 
where the prime denotes an ordinary derivative with respect to the variable 
$v=x^3-x^4$.

\subsubsection{Diagonal case}
Since the field equations for a perfect fluid are still so complicated we will
make a further assumption that will bring the metric into  diagonal form;
namely:  the Killing vector
 $\xi$ being  hypersurface orthogonal.

 The
possibilities are now restricted just to diagonal  subcases $(a)$,  since
diagonal subcases $(b)$ do always admit a further Killing vector tangent to the
Killing orbits $V_2$.

Computing the Einstein tensor for those metrics, one has
\be
G_{14}=0 \quad {\rm and}\quad  G_{24}\not=0 \ .\label{n-40}
\ee
 Consequently, we have chosen a coordinate
chart in such a way that the fluid flow velocity always lies in the two plane
spanned by $\partial /\partial x^2$ and $\partial / \partial x^4$ at each point.
Therefore, we will  have in all cases
\be
u=u_2dx^2+u_4dx^4 \ , \quad {(u_2)^2\over g_{22}}+{(u_4)^2\over g_{44}}=-1
\ . \label{n-42}
\ee

A careful analysis of all the possibilities reveals that, in most cases, there
exist further KVs and, in the instance of null homothetic orbits, the energy 
conditions cannot be fulfilled. Apart from these cases, it is worth mentioning
that whenever $X$ is orthogonal to $u$, the metric and field equations are  
\be
ds^2=e^{2x^3}\left\{ a^2(x^4)e^{-2x^2}\left(dx^1\right)^2- \epsilon b^2(x^4)
\left(dx^2\right)^2+ \left(dx^3\right)^2+\epsilon \left(dx^4\right)^2
\right\}\ ,\label{n-57}\ee
\bea
0&=& 2\epsilon -{1\over b^2}+{a'b'\over ab}+{a''\over a}\ ,\label{n-58}\\
0&=&\left[ {a'\over a}-{b'\over b}\right]^2-\left[ {b''\over b}-{a''\over
a}\right] \left[  -{1\over b^2}+{a'b'\over ab}-{b''\over b}\right]
\ ,\label{n-59}\eea
 a prime indicating a derivative with respect to $(x^4)$, and then one
necessarily has
\be
p=\mu=e^{-2x^3}\left\{ 1+\epsilon {b''\over b} \right\}\ , \label{n-61}
\ee
without previously assuming the existence of a barotropic equation  of state.

\section*{Acknowledgments}
The authors would like to thank Prof. A.A. Coley (Dalhousie University) for
many helpful discussions.

Financial support from DGICYT Research Project No. PB 94-1177 is also
acknowledged.


\begin{thebibliography}{99}
\bibitem{Kramer} D. Kramer, H. Stephani, M.A.H. MacCallum and E. Herlt.
 {\em Exact Solutions of Einstein's Field Equations}. Deutscher Verlag
 der Wissenschaften, Berlin (1980).
\bibitem{Hall88} G.S. Hall, Gen. Rel. Grav., {\bf 20} (1988) 671.
\bibitem{Hall}  G.S. Hall, preprint (1988).
\bibitem{Hall88b} G.S. Hall  in {\em Relativity Today}, Proc.
2nd
Hungarian Relativity Workshop 1987, Ed. Z. Perjes, Singapore,
World Scientific 1988.
\bibitem{Hall88c} G.S. Hall,  Class. Quantum Grav., {\bf 5} (1988)
L77.
\bibitem{Hall90b} G.S. Hall, J. Math. Phys. {\bf 31} (1990) 1198.
\bibitem{HallCos88} G.S. Hall and J. da Costa, J. Math. Phys.,
 {\bf 29} (1988) 2465.
\bibitem{Hall90} G.S. Hall and J.D. Steele, Gen. Rel. Grav., {\bf 22}
 (1990) 457.
\bibitem{CaT} M.E. Cahill and A.H. Taub, Comm. Math. Phys. {\bf 21} (1971) 1.
\bibitem{Eardley} D.M. Eardley, Commun Math. Phys., {\bf 37} (1974) 287.
\bibitem{Eardley2} D.M. Eardley, Phys. Rev. Lett. {\bf 33} (1974) 442.
\bibitem{p1} A.A. Coley, preprint (1995).
\bibitem{p2} B.J. Carr and A.A. Coley, preprint (1996).
\bibitem{p3} B.J. Carr, preprint (1992).
\bibitem{Wu} Wu Z. C. , Gen. Rel. Grav., {\bf 13} (1981) 625.
\bibitem{Shikin} I.S. Shikin,  Gen. Rel. Grav., {\bf 11} (1979) 433.
\bibitem{Palais57} R.S. Palais, Mem. Am. Math. Soc.(1957) no. 2.
\bibitem{Brickell70} F. Brickell and R.S. Clark, {\em Differentiable
 Manifolds}, Van Nostrand (1970)
\bibitem{Bilyalov} R.F. Bilyalov, Sov. Phys., {\bf 8} (1964) 878.
\bibitem{Alex85} D. Alexeevski, Ann. Global. Anal. Geom., {\bf 3} (1985)
 p.59.
\bibitem{Beem} J.K. Beem, Lett. Math. Phys., {\bf 2} (1978) 317.

\bibitem{Wainwright} J. Wainwright, {\it Self-similar solutions of
Einstein's equations} in {\em Galaxies, Axisymmetric systems and relativity}
 , ed. M.A.H. MacCallum, Cambridge (1985) C.U.P.
\bibitem{Barrow86} J.D. Barrow and D.H. Sonoda, Phys. Rep. {\bf 139} (1986) 1.
\bibitem{McInt76} C.B.G. McIntosh, Gen. Rel. Grav., {\bf 7} (1976) 199.
\bibitem{McInt78} C.B.G. McIntosh, Phys. Lett. A {\bf 69} (1978) 1.
\bibitem{McInt79} C.B.G. McIntosh, Gen. Rel. Grav., {\bf 10} (1979) 61.
\bibitem{Ellis67} G.F.R. Ellis, J. Math. Phys., {\bf 8} (1967) 1171.
\bibitem{Stewart68} J.M. Stewart and G.F.R. Ellis, J. Math. Phys.
 {\bf 9} (1968) 1072.
\bibitem{Rosq} K. Rosquist and R.T. Jantzen, Class. Quantum Grav., {\bf 2}
 (1985)  L129.
\bibitem{Ryan} M.P. Ryan and L.C. Shepley, {\em Homogeneous Relativistic
 Cosmologies}, Princeton Univ. Press. (1975)
\bibitem{Barnes79} A. Barnes, J. Phys., {\bf A12} (1979) 1493.
\bibitem{Goenner70} H. Goenner and J. Stachel, J. Math. Phys., {\bf 11}
 (1970) 3358.
\bibitem{Ali1} J. Carot, L. Mas and A.M. Sintes, J. Math. Phys., {\bf 35}
 (1994) 3560.
\bibitem{Uggla92} C. Uggla, Class. Quantum Grav., {\bf 209} (1992) 2287.
\bibitem{Hsu1} L. Hsu and J. Wainwright, Class. Quantum Grav., {\bf 3} (1986)
1105.
\bibitem{Hsu2} J. Wainwright and L. Hsu, Class. Quantum Grav., {\bf 6} (1989)
1409.
\bibitem{Kramer84} D. Kramer,  Class. Quantum
Grav., {\bf 1} (1984) 611.
\bibitem{Kr90} D. Kramer,  Gen. Rel. Grav., {\bf
22} (1990) 1157.  
\bibitem{Mars} M. Mars and J.M.M. Senovilla, Class. Quantum Grav., {\bf
10} (1993) 1633.
\bibitem{Senovilla92}  J.M.M. Senovilla, Class. Quantum
Grav., {\bf 9} (1992) L167.
\bibitem{Hewitt88} C.G. Hewitt, J. Wainwright and S.W. Goode, Class.
 Quantum Grav., {\bf 5} (1988) 1313.
\bibitem{Hewitt91} C.G. Hewitt, J. Wainwright and M. Glaum, Class. Quantum
 Grav., {\bf 8} (1991) 1505.
\bibitem{Hewitt90} C.G. Hewitt and J. Wainwright, Class. Quantum Grav.,
{\bf 7} (1990) 2295.
\bibitem{Schmidt} B.G. Schmidt, Z. Naturforsch {\bf 22a} (1967)
1351.
\bibitem{Ori2} A. Ori and T. Piran, Phys. Rev. {\bf D42} (1990) 1068.
\bibitem{Yahil} B.J. Carr and A. Yahil, Astroph. J. {\bf 360} (1990) 330.
\bibitem{Patel} R.N. Henriksen and K. Patel, Gen. Rel. Grav. {\bf
23} (1991) 527.
\bibitem{Fog} T. Foglizzo and R.N. Henriksen, Phys. Rev.
{\bf D48} (1993) 4645.
\bibitem{Kundt61} W. Kundt, Z. Phys.{\bf 163} (1961) 77.
\bibitem{Ruiz} E. Ruiz and J.M.M. Senovilla, Phys. Rev. D{\bf
45} (1992) 1995.
\bibitem{Agnew} A.F. Agnew and  S.W. Goode, Class. Quantum. Grav., {\bf 11}
(1994) 1725. 
\bibitem{Bergh92} N. Van den Bergh and J. Skea,  Class.
Quantum Grav., {\bf 9} (1992) 527.
\bibitem{Berg91} N. Van den Bergh, P. Wils and M. Castagnino, Class. Quantum.
Grav., {\bf 8} (1991) 947.
\bibitem{Wils91} P. Wils, Class. Quantum. Grav., {\bf
8} (1991) 361. 
\bibitem{Wainwright2} J. Wainwright, J. Phys. A, {\bf 14} (1981) 1131.
\bibitem{Luminet} J.P. Luminet, Gen. Rel. Grav., {\bf 9} (1978) 673.
\bibitem{Hawking73} S.W. Hawking and G.F.R. Ellis, {\em The Large
Scale Structure of Space-time}, Cambridge Univ. Press, Cambridge (1973).
\bibitem{Wainwright79} J. Wainwright, W.C.W. Ince and B.J.
Marshman, Gen. Rel. Grav., {\bf 10} (1979) 259. 

\bibitem{Bergh88} N. Van den Bergh, Class. Quantum. Grav., {\bf 5} (1988)
861. 
 
\end{thebibliography}
\end{document}